# Deposition of SiO$_x$ films by means of atmospheric pressure microplasma jets

J. Benedikt, R. Reuter, D. Ellerweg, K. Rügner, A. von Keudell
Research Department Plasmas with Complex Interactions, Ruhr-University Bochum,
Universitaetsstr. 150, Bochum 44780, Germany

e-mail: jan.benedikt@rub.de

**Abstract**

Atmospheric pressure plasma jet sources are currently in the focus of many researchers for their promising applications in medical industry (e.g. treatment of living tissues), surface modification or material etching or synthesis. Here we report on the study of fundamental principles of deposition of SiO$_x$ films from microplasma jets with admixture of hexamethyldisiloxane [(CH$_3$)$_3$SiOSi(CH$_3$)$_3$, HMDSO] molecules and oxygen. The properties of the deposited films, the composition of the plasma as measured by molecular beam mass spectrometry and the effect of additional treatment of grown film by oxygen or hydrogen atoms will be presented.

**Introduction**

Atmospheric pressure microplasma discharges have attracted a great deal of attention due to their interesting properties such us high electron density inaccessible for typical low pressure plasmas ($n_e$ up to $10^{21}$ m$^{-3}$ and very low gas temperature ($T_g$ as small as 300 K). Moreover, the operation at atmospheric pressure and low power consumption (~ 1 W) make them attractive for on chip implementation and mass production. Up to now, a number of different configurations have been described in the literature for microplasmas operated at atmospheric pressure. Examples are microhollow cathode discharges, capillary plasma electrode discharges, cylindrical dielectric barrier discharges, miniature inductively and capacitively coupled plasmas, jets or devices fabricated in semiconductor structures [1-3]. Microplasma sources can be DC, RF or MW driven and they can be used for example as photonic devices [4,5], as analytical tools [6], for bacterial inactivation [7], for etching of silicon wafers[8] or for localized deposition [9,10].

Here we present the study of the SiO$_x$ thin film deposition by means of atmospheric pressure microplasma jets. We use two different sources: i) a source with coaxial electrode structure with a 250 μm gap and with dielectric barrier at outer electrode and ii) a parallel metal-electrode jet with electrode gap of 1 mm. Both are used for the study of atmospheric pressure deposition of SiO$_x$ thin films. The former one can be operated for a very long times and produces good quality SiO$_2$ films [11,12], whereas the latter source is effective in dissociation of HMDSO, but his operation time is reduced due to deposition on its electrodes.

**Experimental Setup**

Fig. 1 shows a scheme of the microplasma jet. The detailed description of the setup has been published previously [13-15]. A stainless steel capillary tube is inserted into a ceramic tube leaving an annular gap of 250 μm between the tubes. The capillary ends 2 mm prior to the end of the ceramic tube. Around the ceramic tube and 1 mm apart from its end, an aluminum tube serves as counter electrode. A 13.56 MHz radio frequency power supply is attached through a matching network to the capillary or to the outer electrode. Outer electrode is powered in this case. Plasma forming gas (He or Ar in this case) is introduced into the annular space between the ceramic tube and the capillary with a flow rate of 3000 sccm. Additional flow of 160 sccm is guided through the capillary in order to maintain similar gas velocities in the annular space between electrodes and in the capillary. This is especially important when some reactive gas is added to the capillary flow [15]. The reactive gases are in this case HMDSO and oxygen. The HMDSO flow, since it is liquid under standard conditions, is obtained by guiding Ar or He gas through a bubbler containing HMDSO. The plasma can be ignite easily in He by applying root mean square (RMS) voltage of about 100 V. The ignition in Ar is not possible without an external high voltage pulse and therefore an Ar-plasma operation is realized through ignition in He and switching gas flows to Ar. Previous studies assessed the effect of environmental air around the jet on the plasma [13,14]. Nitrogen and oxygen of the surrounding atmosphere do not interfere with the plasma jet, since the main flow through the ceramic tube acts as a barrier.

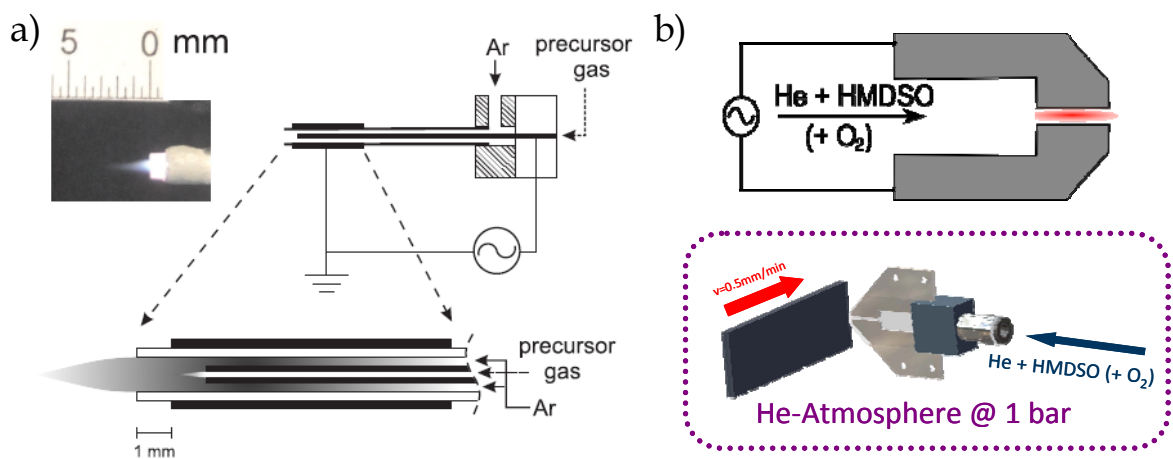

*Fig. 1: The schemes of the microplasma jets used in this work: a) coaxial microplasma jet, b) planar microscale atmospheric pressure plasma jet (µ-APPJ)*

The second source, microscale atmospheric pressure plasma jet (µ-APPJ), is a capacitively coupled microplasma jet consisting of two stainless steel electrodes (length 10 mm, width 1 mm) with a 1 mm gap in between (cf. Fig. 2). The plasma is ignited in this gap filling a volume of 1 x 1 x 10 mm$^3$. The plasma volume is confined on two sides by glass plates. One electrode is connected to a power supply (13.56 MHz, <30 W) through a matching network and the other one is grounded. The gas flow through the µ-APPJ is 5 slm helium with a small admixture of molecular oxygen and HMDSO. The deposition takes place on an moving substrate (v = 0.5 mm/min) to deposit film on a larger area and it takes place in an reactor with helium atmosphere to avoid admixture of surrounding air into the plasma. It has been shown that this microplasma jet is a typical α mode discharge [16].

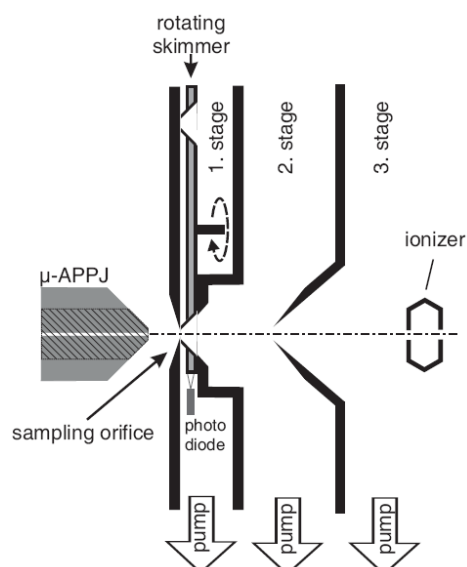

*Fig. 2: The scheme of the sampling system used for the molecular beam mass spectrometry.*

The effluent of this µ-APPJ is analyzed by the molecular beam mass spectrometry (MBMS) system also shown in Fig 2. This MBMS system uses a novel chopper of the molecular beam, which reduces significantly the background pressure in the mass spectrometer and which allows measurements of species concentrations at ppm level. The detailed description can be found elsewhere [17,18].

**Results and discussion (Times 11, bold)**

*Deposition of SiO$_x$ films*

First films deposited from "pure" HMDSO by means of coaxial jet are discussed. By using oxygen containing precursor such as HMDSO, the SiOx films could be theoretically deposited even without addition of oxygen. We have observed this for deposition with the coaxial jet as shown in Fig. 3a), where FTIR spectra of films deposited at varying HMDSO flow rates and without addition of oxygen are show.

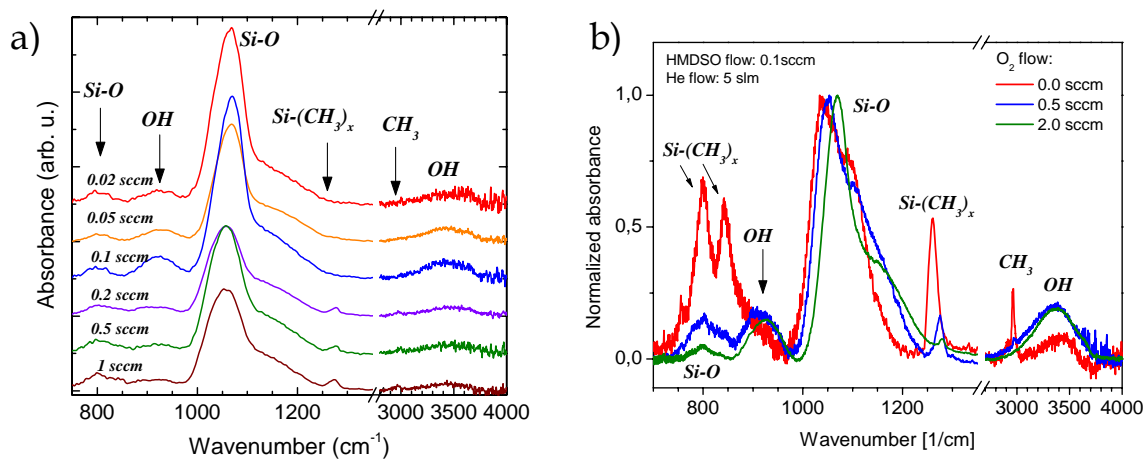

*Fig. 3: FTIR spectra of deposited films. a) Coaxial jet used with different HMDSO flows and without addition of O$_2$, adopted from [12], b) µ-APPJ used.*

At high HMDSO flow rates, the absorption peaks corresponding to the Si-(CH$_3$)$_x$ bending mode at ~1275 cm$^{-1}$ and the CH$_3$ stretching mode at around 2960 cm$^{-1}$ are visible. By decreasing the HMDSO flow rates, the intensities of these carbon related absorption peaks decrease. For very low HMDSO flow rates (below 0.1 sccm), absorption peaks originating from carbon completely disappear: the film changes abruptly from an organic to an inorganic silicon oxide composition. This is surprising, since carbon is usually observed in films deposited from pure HMDSO either at low [19,20] or at atmospheric pressure [21,22]. When O$_2$ flow in the range from 1 to 50 sccm is added to the gas, inorganic SiO$_2$ films without carbon are deposited, which was also corroborated by XPS measurements [12]. These films are still porous with high content of OH groups. The porosity is significantly reduced by heating the substrate to 200 °C.

When µ-APPJ is used and no O$_2$ is added into the gas mixture, a carbon rich film is deposited as shown in Fig. 3b). The SiO$_x$ film with low carbon content is only observed when 2 sccm or more of O$_2$ are added into the plasma. Both plasmas behave quite differently indicating, that Ar plasma chemistry and He plasma chemistry are different.

*Mass spectrometry analysis*

The deposition mechanism of these films, especially under conditions where no oxygen is actively added to the gas flows, is still not well understood. We have proposed that ion-induced polymerization of HMDSO can be responsible for the observed behavior [12], however direct measurement of the depositing species is still not available. To resolve this issue, we have decided to apply MBMS and analyze the composition of the gas mixture emanating from the jets. The µ-APPJ with He as a carrier gas is studied only, because the presence of He significantly enhances the MS signal and improves the detection limit. It is a critical issue in this case because better quality SiO$_x$ films are deposited when HMDSO flow is small, below 100 ppm. The concentrations of stable products of plasma chemistry are similar or lower than this value. Radical densities are most probably below the detection limit of our MBMS system (~ 1 ppm).

The presence of organosilicone compounds detected in atmospheric pressure HMDSO plasmas [23] has been checked by measuring their parent masses under different conditions. An electron energy of 20 eV in the ionizer of the MS is chosen for these measurements to minimize the fragmentation of large molecules. The signal intensities at masses of 28 amu (representative for CO and $N_2$), 73 amu (tetramethylsilan), 75 amu (trimethylsilanol), 131 amu [$(CH_3)_3SiOSi(CH_3)(CH_2)$], 133 amu (pentamethyldisiloxan) and 147 amu (HMDSO) under plasma off, plasma on without addition of $O_2$ and plasma on with addition of $O_2$ conditions are shown in Fig. 4.

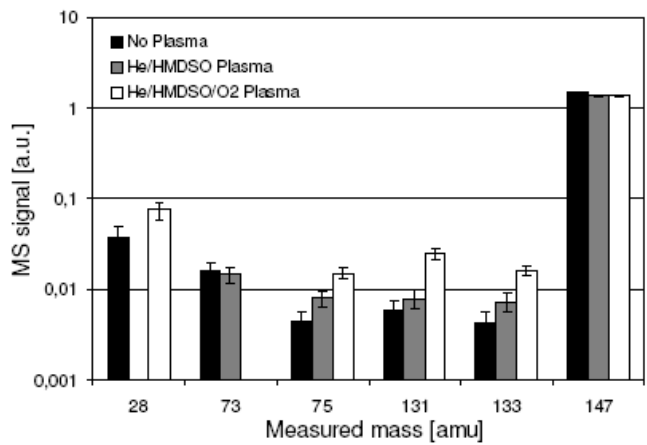

*Fig. 4 The change of the MS signal at selected masses as measured without plasma, with He/HMDSO plasma and He/HMDSO/$O_2$ plasma*

First it can be seen at mass 147, which represents the precursor gas, that the consumption of HMDSO is very low, only 7 % of it are consumed in the plasma, even if $O_2$ is added into the gas mixture. This is typical for atmospheric pressure plasmas and it would corroborate the hypothesis of the ion driven polymerization reaction scheme, which needs large amount of HMDSO molecules for polymerization. The one of the main products of this polymerization scheme is the tetramethylsilan (TMS) detected at mass 73. No change observed at this mass when the plasma is ignited (only the conditions without $O_2$ has been measured) is however in contradiction with this proposed scheme. The main products detected under our conditions are at masses 131 and 133 amu. These are most probably molecules, which are formed directly from HMDSO molecule by a loss of one methyl group. Surprisingly, these molecules are more abundant, when $O_2$ is added to the plasma, even if the HMDSO consumption stays the same. Small increase is also observed at mass 75 amu [trimethilsilanol $(CH_3)_3SiOH$] The larger molecules with molecular mass beyond that of HMDSO (detected usually for example at masses 207 and 221 amu) have densities below our detection limit of ~1 ppm, when no $O_2$ is added, but become measurable when $O_2$ is present in the gas mixture. Slight increase is also observed at mass 28 amu, which is probably CO and indicates that combustion occurs in presence of oxygen.

*Exposure of the surface to atomic oxygen*

The microplasma jet geometry and localization of the deposition/treatment at a spot of few square millimetres allow us to study the role of oxygen in the μ-APPJ deposition process. This is done by alternating the application of He/HMDSO plasma (deposition of the carbon-rich film) and He/$O_2$ plasma (surface oxidation) to the same deposition area, here achieved by a treatment of a rotating substrate by two jets with only He/HMDSO on one hand and He/$O_2$ on the other hand, cf the scheme of the experiment in Fig. 5. Interestingly, the $SiO_x$ film similar to films deposited with HMDSO and $O_2$ in one jet, can be deposited in this way, cf. FTIR spectra in Fig. 5. Moreover, the carbon content is under the detection limit of the FTIR. Apparently, the oxidation of carbon from the surface of the growing film, most probably by atomic oxygen produced in the second jet, is the main carbon elimination step during the deposition of carbon free $SiO_x$ films in this case. It should be noted that application of just $O_2$ flow without plasma in the second jet was not enough to induce this effect and just carbon rich films with very similar FTIR spectra as those deposited without rotating disk and without addition of $O_2$ has been deposited.

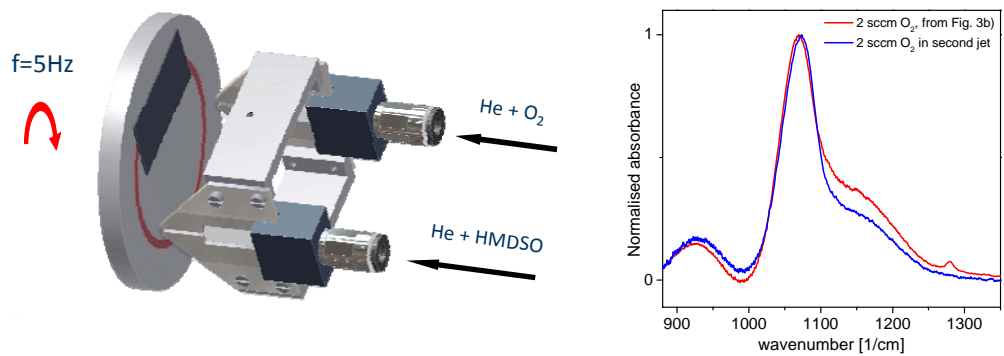

*Fig. 5: The scheme of the experiment with the rotating substrate and the corresponding FTIR spectrum of the deposited film.*

The study of HMDSO plasma chemistry in atmospheric pressure non-equilibrium plasmas is an ongoing process and further analysis are still needed for its better understanding and understanding of the $SiO_x$ film growth, especially regarding the difference between plasmas with He or Ar as a plasma forming gas.

**Acknowledgement**
This work has been done within the project C1 of the research group FOR1123 approved by the German Research Foundation (DFG). This support is gratefully acknowledged.